\newif\ifnote %make
\notetrue    %this makes the draft switch to \notetrue to make it a LCD note

\newif\ifjhep %make
\jhepfalse    %this makes the draft a jhep paper
\newif\ifprep %make
\prepfalse    %this makes the draft a jhep paper
\ifnote              
\documentclass[11pt,a4paper,dvips]{scrartcl}
\usepackage{cite,mcite}
\usepackage{graphicx}
\usepackage{subfig}
\usepackage{lcd}
\usepackage{lcd_title,ifthen}
\usepackage{mathptmx}
\usepackage{helvet}
\usepackage{amsmath}
\usepackage{color}
\usepackage{units}
\usepackage{url}
\usepackage{caption}
\usepackage[margin=1.11in]{geometry}

\makeatletter
\def\url@myurlfontstyle{%
  \@ifundefined{selectfont}{\def\UrlFont{\sf}}{\def\UrlFont{\small\ttfamily}}}
\makeatother
\long\def\symbolfootnote[#1]#2{\begingroup%
\def\thefootnote{\fnsymbol{footnote}}\footnote[#1]{#2}\endgroup}
 \lcdnote{2012-009}
\newlength{\capindent}
\newlength{\capwidth}
\newlength{\figwidth}
\newcommand{\icaption}[2][!*!,!]{\hspace*{\capindent}%
  \begin{minipage}{\capwidth}
    \ifthenelse{\equal{#1}{!*!,!}}%
      {\caption{#2}}%
      {\caption[#1]{#2}}
      \vspace*{3mm}
  \end{minipage}}
\fi
\ifprep
\documentclass[11pt,a4paper,dvips]{scrartcl}    
\usepackage{cite,mcite}
\usepackage{graphicx}
\usepackage{subfig}
\usepackage{lcd}
\usepackage{lcd_title,ifthen}
\usepackage{mathptmx}
\usepackage{helvet}
\usepackage{amsmath}
\usepackage{color}
\usepackage{units}
\usepackage{url}
\usepackage{caption}
\makeatletter
\def\url@myurlfontstyle{%
  \@ifundefined{selectfont}{\def\UrlFont{\sf}}{\def\UrlFont{\small\ttfamily}}}
\makeatother
\long\def\symbolfootnote[#1]#2{\begingroup%
\def\thefootnote{\fnsymbol{footnote}}\footnote[#1]{#2}\endgroup}
\preprint{CERN-PH/2012-XXX}
\newlength{\capindent}
\newlength{\capwidth}
\newlength{\figwidth}
\newcommand{\icaption}[2][!*!,!]{\hspace*{\capindent}%
  \begin{minipage}{\capwidth}
    \ifthenelse{\equal{#1}{!*!,!}}%
      {\caption{#2}}%
      {\caption[#1]{#2}}
      \vspace*{3mm}
  \end{minipage}}
\fi
\ifjhep              
\documentclass[11pt,a4paper]{article}
\usepackage{jheppub}
\usepackage{subfig}
\usepackage{ifthen}
\usepackage{units}
\fi

\def\Tr{{\rm Tr}}
\def\mafz{{\rm AFZ}'}
\def\beq{\begin{equation}}
\def\eeq{\end{equation}}
\def\tev{\, {\rm TeV}}

\def\mzp{M_{Z'}}

\def\xtitle{Physics performances for $Z'$ searches at \\ $\sqrt{s}$=3 TeV~and~1.4~TeV
at CLIC
}

\def\xabstract{Extra neutral gauge bosons ($Z'$) are predicted in many extensions of the Standard Model (SM). 
In the minimal anomaly-free $Z'$ model ($\mafz$), the phenomenology is controlled by only three parameters 
beyond the SM ones, the $Z'$ mass and two effective coupling constants $g'_Y$ and $g'_{BL}$. 
We study the $Z'$ $5\sigma$ discovery potential in $e^+e^-$ collisions at $1.4\tev$ and $3\tev$ at CLIC.  
Assuming LHC discovers a $Z'$ of $5\tev$ mass, the expected accuracies on the $Z'\mu^+\mu^-$ couplings are presented. 
We discuss also the requirements on detector performance and beam polarization.}

\def\ee{\ensuremath{e^+ e^-}}%
\ifjhep
\title{\xtitle}
\author[a]{Jean-Jacques Blaising,}
\author[b]{James D. Wells}
\affiliation[a]{Laboratoire d'Annecy-le-Vieux de Physique des Particules, Annecy-le-Vieux, France}
\affiliation[a]{CERN, Theory Group, Geneva, Switzerland}

\emailAdd{Jean-Jacques.Blaising@lapp.in2p3.fr}
\emailAdd{James.Wells@cern.ch}
\abstract{\xabstract}
\notoc
\fi
\begin{document}
\ifjhep
\message{**:jhep maketitle}
%%\date{\currenttime}
\date{July 3, 2012}
\maketitle
\flushbottom
\fi

\ifnote
\message{**:start: note title page}
\begin{titlepage}
%
% Define logo file
%\logo{LCD_logo.eps}
%\logo{../CLIC_Logo.epsi}
% As long as we do not have a logo leave some space instead
%\vskip 35mm
%\logob{../CERN_Logo.epsi}
% As long as we do not have a logo leave some space instead
\vskip 35mm
%
% Print version number of document
\mydocversion
%
% Title of the paper (JJ)
%\title{Determination of Z' $\mu^+\mu^-$ couplings at CLIC}
\title{\xtitle}
%
% Author(s) of the paper (JJ)
\author{
Jean-Jacques Blaising\affiliated{1},
James D. Wells\affiliated{2},
%%             Initials. Name\affiliated{1}
             }
% Affiliations
\affiliation[3]{LAPP, Laboratoire d'Annecy-le-Vieux de Physique des Particules, Annecy-le-Vieux, France}, \newline
\affiliations{\affiliation[2]{CERN, Geneva, Switzerland}
%\affiliation[2]{University of California at Santa Cruz, Santa Cruz, CA, USA}, \newline
%\affiliation[4]{University of Cambridge, Cambridge, UK} 
}
%
% Give date: \today for drafts, a fixed date for final papers
\date{July 3, 2012}
%%\date{August 12, 2011}
%\date{May 7, 2012}
%
% The abstract  (JJ)
\begin{abstract}
\xabstract
\end{abstract}
%
% Give name, place and date of the conference if presented
%% (JJ next 2 line commented out)
%\presented{Name of Conference, Geneva, January 1, 2009}
% Give name of the journal if submitted
%\submitted{Name of Journal}
%
\end{titlepage}
\fi
\ifprep
\message{**:start: prep title page}
\begin{titlepage}
% Define logo file
%\logo{LCD_logo.eps}
\logo{../CLIC_Logo.epsi}
\logob{../CERN_Logo.epsi}
% As long as we do not have a logo leave some space instead
\vskip 35mm
%
% Print version number of document
\mydocversion
%
% Title of the paper
\title{
Physics performances for $Z'$ searches at CLIC
}
% Author(s) of the paper (JJ)
\author{
J-J. Blaising\affiliated{3},
J. Wells\affiliated{1},
%%             Initials. Name\affiliated{1}
             }
% Affiliations
\affiliations{\affiliation[1]{CERN, Geneva, Switzerland}, \newline
\affiliation[3]{LAPP, Laboratoire d'Annecy-le-Vieux de Physique des Particules, Annecy-le-Vieux, France} \newline
}
%
% Give date: \today for drafts, a fixed date for final papers
\date{\today}
%\date{May 7, 2012}
% The abstract  (JJ)
\begin{abstract}
\xabstract
\end{abstract}
% Give name of the journal if submitted
%\submitted{JHEP}
\end{titlepage}
\end{document}
\fi

\ifjhep
\tableofcontents
\fi

\clearpage

\section{Minimal Anomaly-Free $Z'$ Models}

Many extensions of the Standard Model (SM) call for more gauge forces beyond the 
ordinary $SU(3)_c\times SU(2)_L\times U(1)_Y$. For example, a higher rank gauge group may spontaneously break down to the SM plus additional gauge groups. Or, some braneworld constructions of gauge interactions may necessarily imply the existence of other gauge interactions beyond the SM ones. In such cases, 
the simplest gauge extension beyond the SM is an additional abelian $U(1)$ symmetry.

For self-consistency of quantum field theory it is generally necessary that the gauge forces have no anomalies. 
These include the pure gauge anomalies, such as $\Tr [U(1)'^3]$, the mixed anomalies such 
as $\Tr [ U(1)'SU(N)^2]$, and the gauge-gravity-gravity anomalies such as $\Tr [ U(1)']$. 
It is often the case that a new $U(1)'$ will need additional fermions in the spectrum to cancel the anomalies. 
These exotic fermions will get mass associated with the new $U(1)'$ breaking scale, and thus $Z'$ bosons 
and exotic fermions have masses within close proximity to each other. 
It is a detailed model building question in that case whether the fermions are lighter or heavier than  the gauge bosons, 
and then a detailed phenomenological analysis to determine which one would be seen first at a high-energy 
collider.

On the other hand, there are models, which we call `Minimal Anomaly-Free $Z'$' models, or $\mafz$ for short, 
which are anomaly free with respect to the SM gauge groups and particle content alone. There are no additional fermions that are necessary in the spectrum, and indeed if 
there were, their charges would have to conspire to keep all anomalies zero. This is trivially satisfied if 
exotic fermions are vector-like, in which case they can have direct mass terms without the need of 
the $U(1)'$ breaking, and so are likely to be at a mass scale much heavier than the $U(1)'$ breaking scale that 
gives mass to the $Z'$.  

The simplifying nature of the theory and phenomenology of the $\mafz$ model~\cite{Salvioni:2009mt} is attractive 
for our purposes of demonstrating without complications the intrinsic value of a high-energy $e^+e^-$ collider 
to discover and study the effects of a new $Z'$ gauge boson. Since an anomaly free $U(1)'$ with respect to 
the SM spectrum must necessarily be a linear combination of hypercharge 
and $B-L$ (baryon number minus lepton number),
\beq
Q_f=g'_YY_f+g'_{BL}(B-L)_f,
\eeq
the phenomenology of these models is determined by just three parameters, $g'_Y$, $g'_{BL}$ and $M_{Z'}$. 
Note, kinetic mixing of the $U(1)'$ with hypercharge can be diagonalized away, having only the effect 
of changing the values of $g'_Y$ and $g'_{BL}$.  This sets up an excellent example theory with few parameters to 
investigate $Z'$ capabilities at $e^+e^-$ colliders~\cite{OtherZprimestudies}.

Regarding the collider phenomenology, below the $Z'$ peak, the $Z'$ can be detected through precision 
measurements allowing observations of small deviations of observables from their SM predictions.
In this paper we study the $\mafz$ $Z'$ discovery potential in  $e^+e^- $ collisions at $1.4\tev$ and $3\tev$ at CLIC. 
Next, assuming LHC discovers a $Z'$ of $5\tev$ mass, the expected accuracies on the $Z'\mu^+\mu^-$ couplings 
are determined. 
The discovery potential and the couplings determination are based on the measurement of several observables.
In this analysis we use three observables, namely, the total cross-section $\sigma (e^+e^- \to \mu^+\mu^-) $,
the forward-backward asymmetry $A_{FB} $ and the left-right asymmetry $A_{LR} $, with
\begin{eqnarray}
\sigma_{tot}=\sigma_F+\sigma_B,
\hspace{1.1cm}
A_{FB} = \frac{\sigma_F-\sigma_B }{ \sigma_F+\sigma_B },
 \hspace{1.2cm}
 A_{LR} = \frac{\sigma_L-\sigma_R }{ \sigma_L+\sigma_R }.
\end{eqnarray}
The observables $\sigma_{tot}$ and $A_{FB}$ are measured with respect to unpolarized electron and positron beams. The $A_{LR}$ asymmetry is defined with respect to $+80\%$ and $-80\%$ polarized electron beams for $\sigma_L$ and $\sigma_R$ respectively. The positron beam is considered unpolarized.

%%%%%%%%%%%%%%%%%%%%%%%%%%%%%%%
\section{Event Simulation and Selection} \label{dtsmcs}

Performance of high-energy leptons at CLIC have been studied in the framework of the 
$\mathrm{CLIC\_ILD}$ and $\mathrm{CLIC\_SiD}$ detector models and reported in~\cite{Slepton:2012jjb}.
On the basis of this work, it is valid to do
physics performance studies for lepton final state processes at
generator level provided that one takes into account the detector acceptance, the lepton reconstruction 
and identification efficiency. 
The $Z'$ study reported in this paper is performed at generator level.

SM events are generated using
{\sc Whizard~6.4} and $\mafz$ events with {\sc Whizard~2.0}~\cite{Whizard:2008}. 
Beamstrahlung effects on the luminosity spectrum are included using results of the CLIC beam simulation 
for the CDR accelerator parameters~\cite{Braun:2008zzb}. 
The luminosity spectrum is obtained from the 
{\sc GuineaPig}~\cite{c:thesis} beam simulation, which is then used as input 
to WHIZARD, while simultaneously enabling initial state radiation (ISR) and final state radiation (FSR).

In the presence of a $Z'$, the cross-section values of   
$\sigma (e^+e^- \to \gamma /Z /Z' \to  \mu^+\mu^-) $ differ from the SM value by an amount dependent 
upon the $Z'$ mass, and the couplings $g'_Y$  and $g'_{BL}$.
The deviations from the SM are small, especially in the case of a $Z'$ being inaccessible to the LHC, $M_{Z'}\gg \sqrt{s}$.
Therefore radiative effects have to be included such that the theoretical predictions match with the expected experimental precision.

In $e^+e^- $ collisions, photons are radiated through initial state radiation (ISR) or machine beam-strahlung.
When a photon is radiated, the center-of-mass of the interaction is not the nominal one. 
Due to ISR and beam-strahlung effects, the $\sqrt{s}$ spectrum has a long tail
down to very low values. For the process  $e^+e^- \to \mu^+\mu^- $, the $\sqrt{s}$ spectrum has a second peak
at $\sqrt{s}=M_Z$ due to radiative return to the $Z$ resonance. 

Events with such hard photons have less $e^+e^-$ center-of-mass energy available and so are much less 
sensitive to a $Z'$; therefore,  we eliminate them
by cuts on the energy and angles of the outgoing muons.
In addition, other SM processes produce $  \mu^+\mu^- $ final states. The most important of 
these additional contributions are listed in Table~\ref{tab:backg1400}.
At CLIC, beam-induced background $ \mu^+\mu^- $ final state events are produced in the process  
$\ee \rightarrow \gamma \gamma \rightarrow \mu^+ \mu^-$.
Both types of background events, SM and beam-induced, must be suppressed 
to preserve the purity of the $\ee \rightarrow  \mu^+ \mu^-  $ sample.   

\begin{table}[tp]
\begin{center}
\begin{tabular}{|l|l|l|}
\hline
Process                    &$\sigma \times Br $ (fb)         &~~$\sigma \times Br$ (fb) \\ 
%$\sqrt{s}=1.4\tev$    &~~fiducial volume cuts & ~~final selection cuts \\ \hline
$\sqrt{s}=1.4\tev$             &$10^\circ < \theta \mathrm{(\mu^{\pm}) <170^\circ }$ and  $P_T \mathrm{ (\mu^{\pm}) > 5~GeV} $
& ~~final selection cuts   \\\hline
%$\sqrt{S}$                 &~~1.4 TeV                            &~~ 1.4\\\hline
$\ee \rightarrow  \mu^+ \mu^-  $ &~~156  &~~23.6 \\
$\ee \rightarrow  \mu^+ \mu^- \nu_e \nu_e $ &~~44.7 &~~0.002\\
$\ee \rightarrow  \mu^+ \mu^- \nu_{\mu} \nu_{\mu} $ &~~14.5 &~~0.027 \\
$\ee \rightarrow  \mu^+ \mu^- e^+ e^- $ &~~1690 &~~$<$ 0.0001 \\ \hline
%$\ee \rightarrow  W^+ W^- \nu_e \nu_e $ &~~3.24E+00 &~~\leq~1.00E-04 \\
%$\ee \rightarrow  \tau^+ \tau^- $    &~~1.03E+01 &~~<1.0E-04 \\\hline
\end{tabular}
\caption{SM $\ee \rightarrow  \mu^+ \mu^- $  processes, cross sections times branching ration ($\sigma \times Br$) 
with angular and $P_T$ cuts, and with final selection cuts, at 1.4 TeV \label{tab:backg1400}}

\end{center}
\end{table}

The detector angular acceptance is  defined by $10^\circ < \theta \mathrm{(\mu^{\pm}) <170^\circ }$,
$\theta$ is the angle of the $\mu^+$ or the $\mu^-$ with respect to the beam.
In this region the muons are measured with high efficiency and excellent momentum resolution.   
To suppress the  beam-induced background, 
$\ee \rightarrow \gamma \gamma \rightarrow \mu^+ \mu^-$ and    
$\ee \rightarrow \gamma \gamma \rightarrow $ hadrons,    
a cut on $P_T \mathrm{ (\mu^{\pm}) }$ is applied, $P_T \mathrm{ (\mu^{\pm}) > 5~GeV} $,
where $P_T $ is the transverse momentum.
To reduce the hard photon events and the contributions of the SM background processes,
additional cuts are applied:
\begin{itemize}
\item dimuon energy, $\mathrm{E(\mu^+)+E(\mu^-)}> E_{\rm min}$ ,
\item acoplanarity, $0^\circ < \Delta \phi \mathrm{(\mu^+,\mu^-) < 5^\circ }$, 
where $\Delta  \phi \mathrm{(\mu^+,\mu^-)}\equiv |\phi_{\mu^+}-\phi_{\mu^-}-\pi |$
~(that is, $\phi_{\mu^+}$ must be nearly back-to-back to $\phi_{\mu^-}$ in the azimuthal plane)
\item angle of the dimuon missing momentum vector, $ 0 < \theta_{miss}(\mu^+,\mu^-) <5^\circ$ 
~(that is, the missing momentum vector polar angle must be very close to beam)
\end{itemize}
where $\mathrm{E_{min}}=1.2\tev$ for $\mathrm {\sqrt{s}}$=1.4 TeV and $E_{\rm min}=2.5\tev$ 
for $\mathrm {\sqrt{s}}=3\tev$.

The muon reconstruction and identification efficiency is 98\% at $3\tev$; it decreases to 97\% in the presence of the 
beam-induced background from $\gamma \gamma\to {\rm hadrons}$. At $1.4\tev$ it is  99\% and 98\% respectively.   
\noindent Table~\ref{tab:backg1400} and Table~\ref{tab:backg3000} show the cross section $\times$ Br values of the 
dimuon final state processes, without and with the final selection cuts at 1.4 and 3.0 TeV.
With these cuts the backgrounds are reduced to near negligible levels in comparison to the signal. 

\begin{table} [tp]
\begin{center}
\begin{tabular}{|l|l|l|}
\hline
Process                                          &$\sigma \times Br $ (fb)               &~~$\sigma \times Br$ (fb)\\ 
%$\sqrt{s}=3\tev$                              &~~fiducial volume cuts & ~~final selection cuts   \\\hline
$\sqrt{s}=3\tev$             &$10^\circ < \theta \mathrm{(\mu^{\pm}) <170^\circ }$ and  $P_T \mathrm{ (\mu^{\pm}) > 5~GeV} $
& ~~final selection cuts   \\\hline
%$\sqrt{S}$                                       &~~3.0 TeV                            &~~3.0 TeV   \\\hline
$\ee \rightarrow  \mu^+ \mu^-  $ &~~82.3          & ~~4.86  \\
$\ee \rightarrow  \mu^+ \mu^- \nu_e \nu_e $ &~~65.6     & ~~$<$ 0.001 \\
$\ee \rightarrow  \mu^+ \mu^- \nu_{\mu} \nu_{\mu} $ &~~4.4    & ~~0.011\\
$\ee \rightarrow  \mu^+ \mu^- e^+ e^- $ &~~1590    & ~~ $<$ 0.0001  \\ \hline
%$\ee \rightarrow  W^+ W^- \nu_e \nu_e $ &~~1.02E+00     & ~~\leq~1.00E-04  \\
%$\ee \rightarrow  \tau^+ \tau^- $    &~~ 7.28E+00               &~~<1.0E-04 \\\hline
\end{tabular}
\caption{SM $\ee \rightarrow  \mu^+ \mu^- $  processes, cross sections times branching ration ($\sigma \times Br$) 
with angular and $P_T$ cuts, and with final selection cuts, at 3 TeV \label{tab:backg3000}}
\end{center}
\end{table}

%%%%%%%%%%%%%%%%%%%%%%%%%%%%%%%%%%%%
\section{Discovery Potential} \label{dtsmcs again2}

To estimate the $Z'$ discovey potential, the SM predictions of the observables $\sigma(SM)$, $A_{FB}(SM)$ and $A_{LR}(SM)$ as well as
the $\mafz$ predictions of the observables $\sigma(\mafz)$, $A_{FB}(\mafz)$ and $A_{LR}(\mafz)$ are computed for different
values of $M_{Z'}$, $g'_Y$ and $g'_{BL}$.
 For each observable the $\chi^2$ is computed, 
defined as the difference between the SM value and the $\mafz$ value:
\begin{eqnarray}
\resizebox{0.9\textwidth}{!} {
$\chi_{\sigma}^2=\frac{(\sigma(\mathrm{SM})-\sigma(\mathrm{\mafz}))^2}{\Delta\sigma(\mathrm{SM})^2} $,  \hspace{0.25cm}
$\chi_{A_{FB}}^2=\frac{(A_{FB}(\mathrm{SM})-A_{FB}(\mathrm{\mafz}))^2}{ \Delta A_{FB}(\mathrm{SM})^2}$, \hspace{0.25cm}
$\chi_{A_{LR}}^2=\frac{(A_{LR}(\mathrm{SM})-A_{LR}(\mathrm{\mafz}))^2}{\Delta A_{LR}(\mathrm{SM})^2} $
}
\end{eqnarray}
where $  \Delta\sigma(\mathrm{SM}),~ \Delta A_{FB}(\mathrm{SM})~ \mathrm {and}~ \Delta A_{LR}(\mathrm{SM})$ 
are the experimental errors on the measurement of the SM observables. The theory computational errors are negligible in comparison.  

The presence of a $Z'$ induces deviations of these observables from their SM predictions. The quantity
$\chi_{sum}^2=\chi_{\sigma}^2+\chi_{A_{FB}}^2+\chi_{A_{LR}}^2$ is an estimator of the sensitivity
to a $Z'$.
Given that the deviations from the SM are small, systematic errors on detector performance, luminosity and polarization measurement 
must be taken into account. In this study we assume an electron polarization of $\pm 80\%$ 
and the following systematic errors:
\begin{itemize}
\item error on $\sigma$ from  $\mu^{\pm}$ reconstruction and identification efficiency: $\Delta \sigma / \sigma$= 1\% 
\item error on $A_{FB}$ from $\mu^{\pm}$ charge confusion: $\Delta A_{FB} / A_{FB}$= 1\% 
\item error on $\sigma$ from luminosity determination: $\Delta \sigma / \sigma$= 0.5\% 
\item error on $A_{LR}$ from polarization measurement: $\Delta A_{LR} / A_{LR}$= 1\%
\end{itemize}

Under these realistic beam and detector conditions, the 
sensitivity to $\mafz$ model parameters $\mzp$, $g'_Y$ and $g'_{BL}$ are estimated for two
different values of the center-of-mass energy. 
In the first set of figures that we describe below, we fix the 
value of $\mzp$ at $5\tev$ and investigate the sensitivities to discovery using different observables for various values of the 
coupling constants in the $g'_Y$ and $g'_{BL}$ plane.  We then show plots of the sensitivity in the plane for various integrated 
luminosities and various $\mzp$ masses. This is done for a $1.4\tev$ machine. We then show the same sensitivity plots for a $3\tev$ 
machine, but for $\mzp=6\tev$.  Finally, we show a plot of the $\mzp$ mass $5\sigma$ discovery reach as a function 
of integrated luminosity for $1.4\tev$ and $3\tev$ CLIC and for $g'_Y$ and $g'_{BL}$ coupling values.

\begin{figure}[tp]
\begin{center}
\resizebox{\textwidth}{!} {
\begin{tabular}{c}
\hspace{-1.cm}
\subfloat[Total cross-section $\sigma (e^+e^- \to \mu^+\mu^-) $]
{\includegraphics[width=0.49\textwidth,clip]{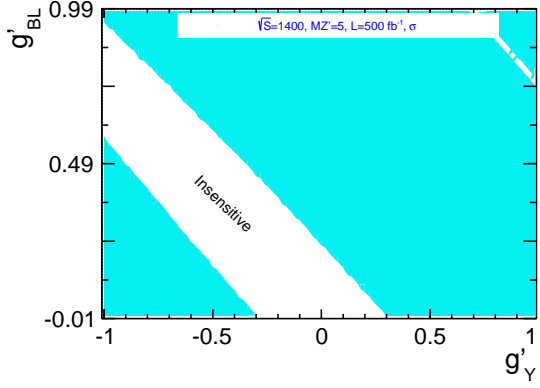}}
\subfloat[Forward-Backward asymmetry]
{\includegraphics[width=0.49\textwidth,clip]{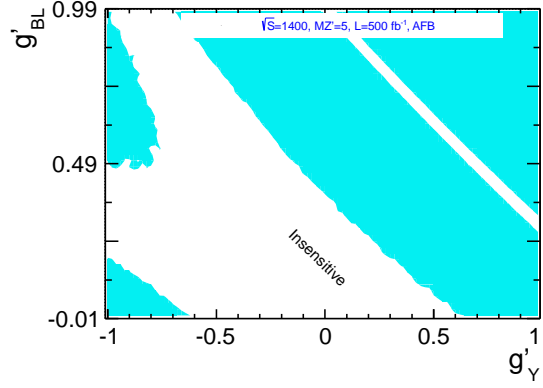} }\\
\subfloat[Left-Right asymmetry]
{\includegraphics[width=0.49\textwidth,clip]{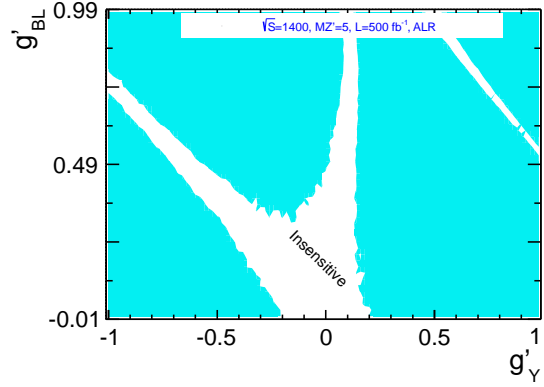} }
\end{tabular}
}
\end{center}
\caption{$5\sigma$ discovery potential in ($g'_Y,g'_{BL}$) plane, $\mzp=5\tev$, L=500 $\mathrm{fb^{-1}}$ and $\sqrt{s}=1.4\tev$,
determined from different observables, (a) total cross-section $\sigma$, (b) forward-backward asymmetry $A_{FB}$, 
and (c) left-right asymmetry $A_{LR}$.}
\label{fig:DISCP1_1400}
\end{figure}

First, Figure~\ref{fig:DISCP1_1400} shows the $5\sigma$ 
discovery potential at $1.4\tev$ in the ($g'_Y,g'_{BL}$) plane 
for $\mzp=5\tev$ and L=500 $\mathrm{fb^{-1}}$
determined from different observables, (a) total cross section $\sigma$, (b) forward-backward asymmetry $A_{FB}$, and
(c) left-right asymmetry $A_{LR}$. The white region corresponds to the region where the $Z'$ cannot be detected. 

\begin{figure}[tp]
\begin{center}
\resizebox{\textwidth}{!} {
\begin{tabular}{c}
\hspace{-1.cm}
\subfloat[$\sigma $ +  $A_{FB}$]
{\includegraphics[width=0.50\textwidth,clip]{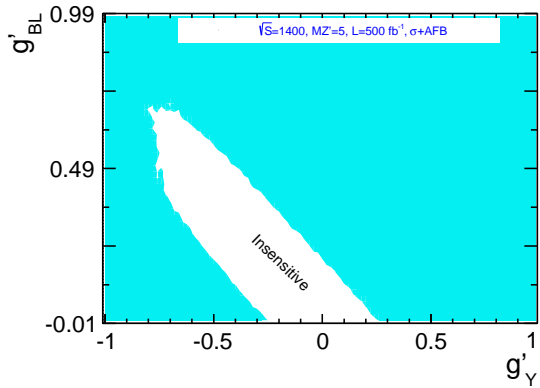}}  
\subfloat[$\sigma $ +  $A_{FB}$ + $A_{FB}$ ]
{\includegraphics[width=0.50\textwidth,clip]{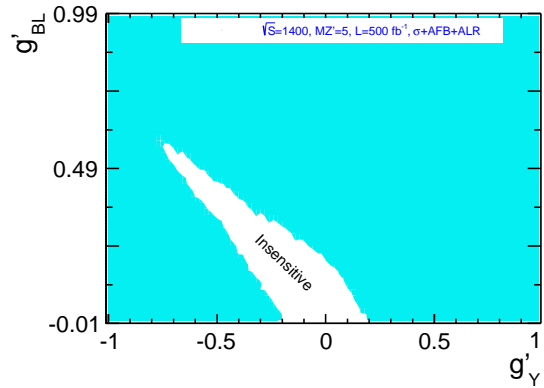} }
\end{tabular}
}
\end{center}
\caption{$5\sigma$ discovery potential in ($g'_Y,g'_{BL}$) plane, 
$\mzp=5\tev$, L=500 $\mathrm{fb^{-1}}$ and $\sqrt{s}=1.4\tev$, determined from combined observables, 
(a)  $\sigma$+ $A_{FB}$, (b)  $\sigma$+$A_{FB}$+$A_{LR}$ .}
\label{fig:DISCPC_1400}
\end{figure}

Figure~\ref{fig:DISCPC_1400} shows the $5\sigma$ discovery potential at $1.4\tev$ in the ($g'_Y,g'_{BL}$) plane 
for $\mzp=5\tev$ and L=500 $\mathrm{fb^{-1}}$ determined from the combined observables, 
(a) $\sigma$ + $A_{FB}$, (b) $\sigma$ + $A_{FB}$+$A_{LR}$.
The observable $A_{LR}$ increases slightly the discovery region for  $g'_Y < 0$.

\begin{figure}[tp]
\begin{center}
\resizebox{\textwidth}{!} {
\begin{tabular}{c}
\hspace{-1.cm}
\subfloat[$\mzp=5\tev$, L=250, 500 and 1000 $\mathrm{fb^{-1}}$ ]
{\includegraphics[width=0.50\textwidth,clip]{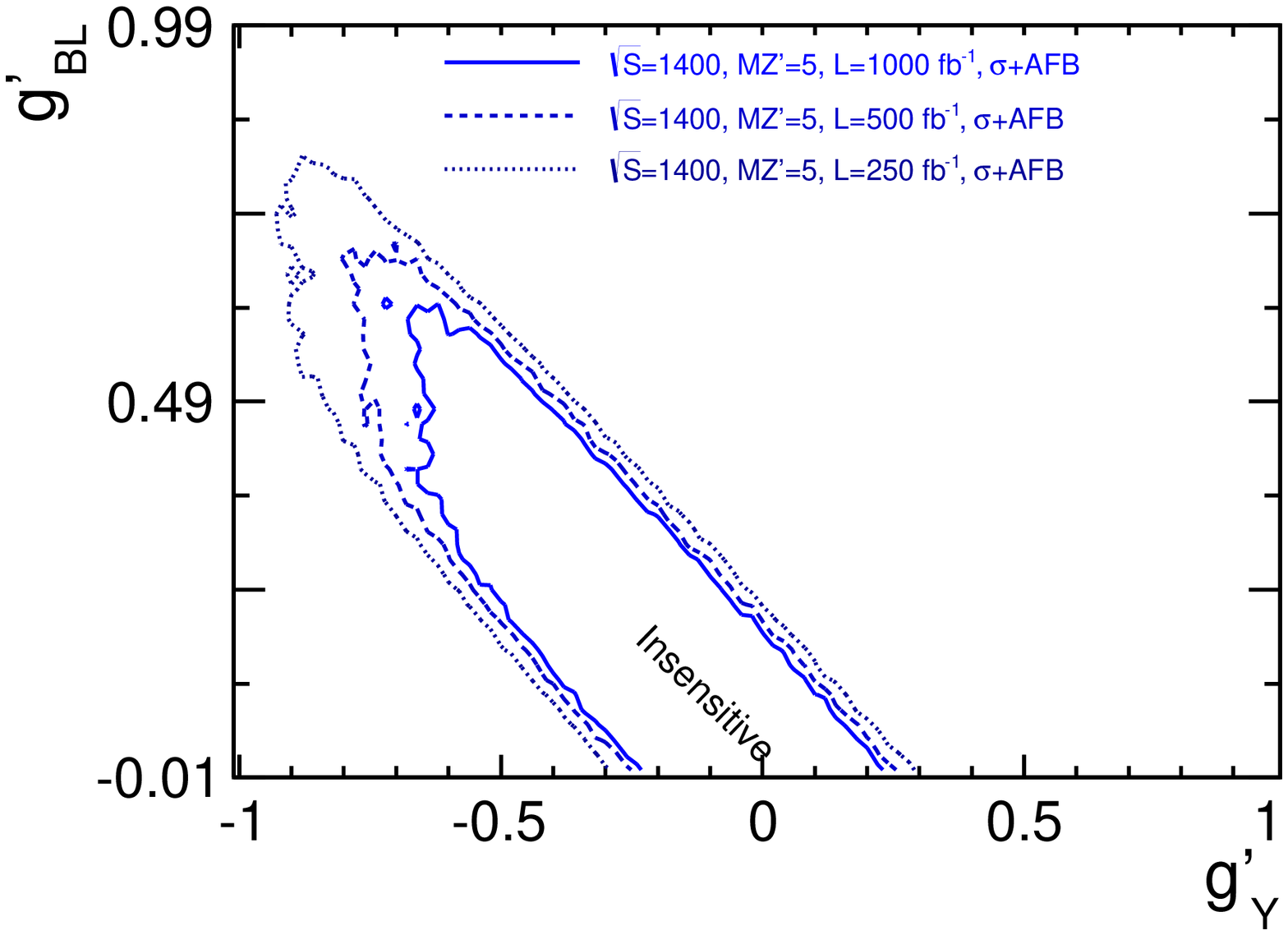}}  
\subfloat[L=500 $\mathrm{fb^{-1}}$  $\mzp=3, 4, 5\tev$]
{\includegraphics[width=0.50\textwidth,clip]{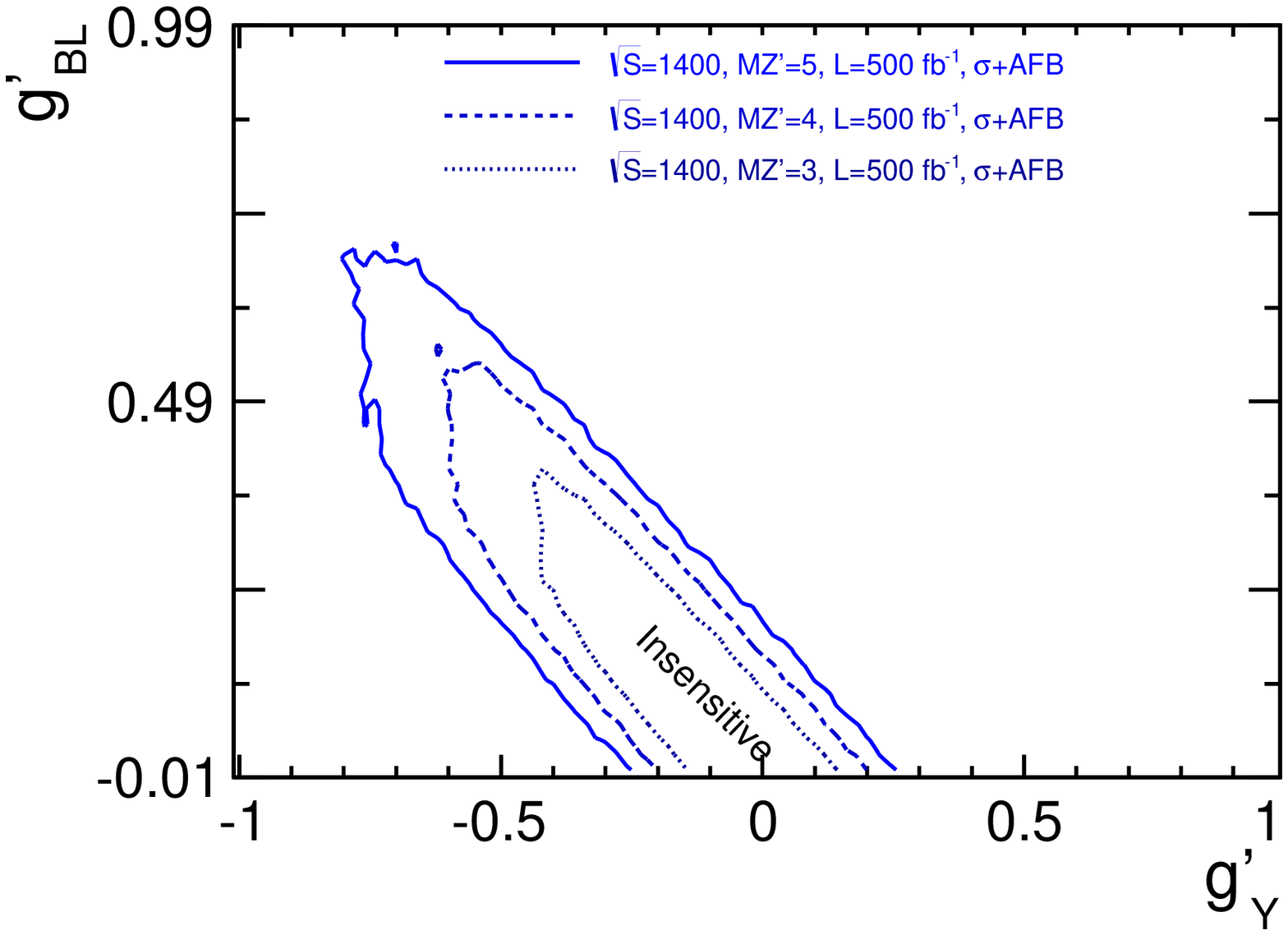} }
\end{tabular}
}
\end{center}
\caption{$5\sigma$ discovery potential in ($g'_Y,g'_{BL}$) plane, determined from combined observables 
$\sigma$+$A_{FB}$ at  $\sqrt{s}=1.4\tev$ for
(a) $\mzp=5\tev$ and different luminosities,
(b) L=500 $\mathrm{fb^{-1}}$ and different $\mzp$ values}.
\label{fig:DISCP3:1400}
\end{figure}

Figure~\ref{fig:DISCP3:1400} shows the $5\sigma$ discovery potential in the ($g'_Y,g'_{BL}$) plane,  
determined from the combined observables $\sigma$ + $A_{FB}$, at $1.4\tev$, 
(a) $\mzp=5\tev$ and different luminosity values, 
(b) L=500 $\mathrm{fb^{-1}}$ and different $\mzp$ values.

\begin{figure}[tp]
\begin{center}
\resizebox{\textwidth}{!} {
\begin{tabular}{c}
\hspace{-1.cm}
\subfloat[$M_{Z'}=5$ TeV, L=250, 500 and 1000 $\mathrm{fb^{-1}}$ ]
{\includegraphics[width=0.50\textwidth,clip]{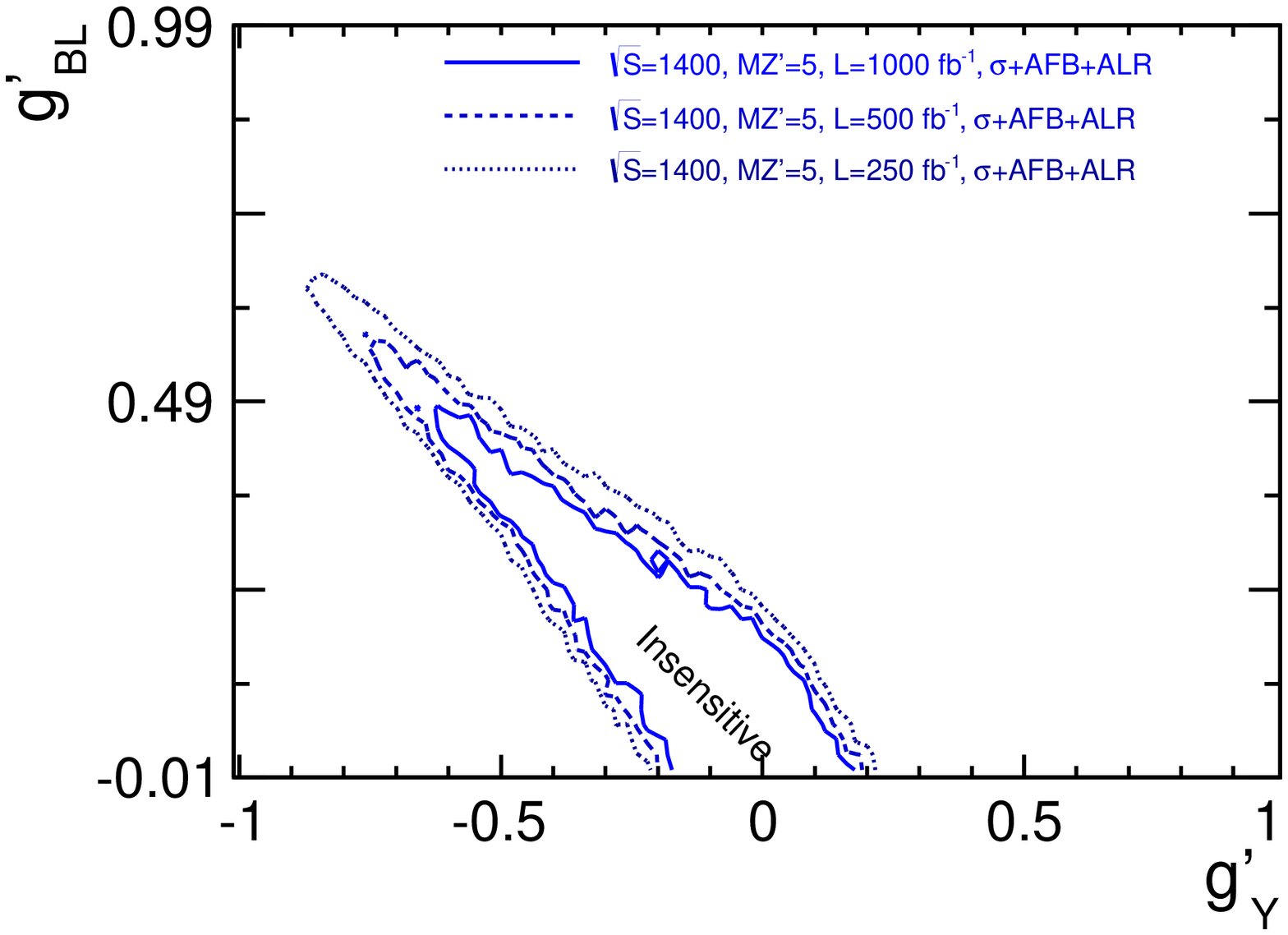}}  
\subfloat[L=500 $\mathrm{fb^{-1}}$  $\mzp=3, 4, 5\tev$]
{\includegraphics[width=0.50\textwidth,clip]{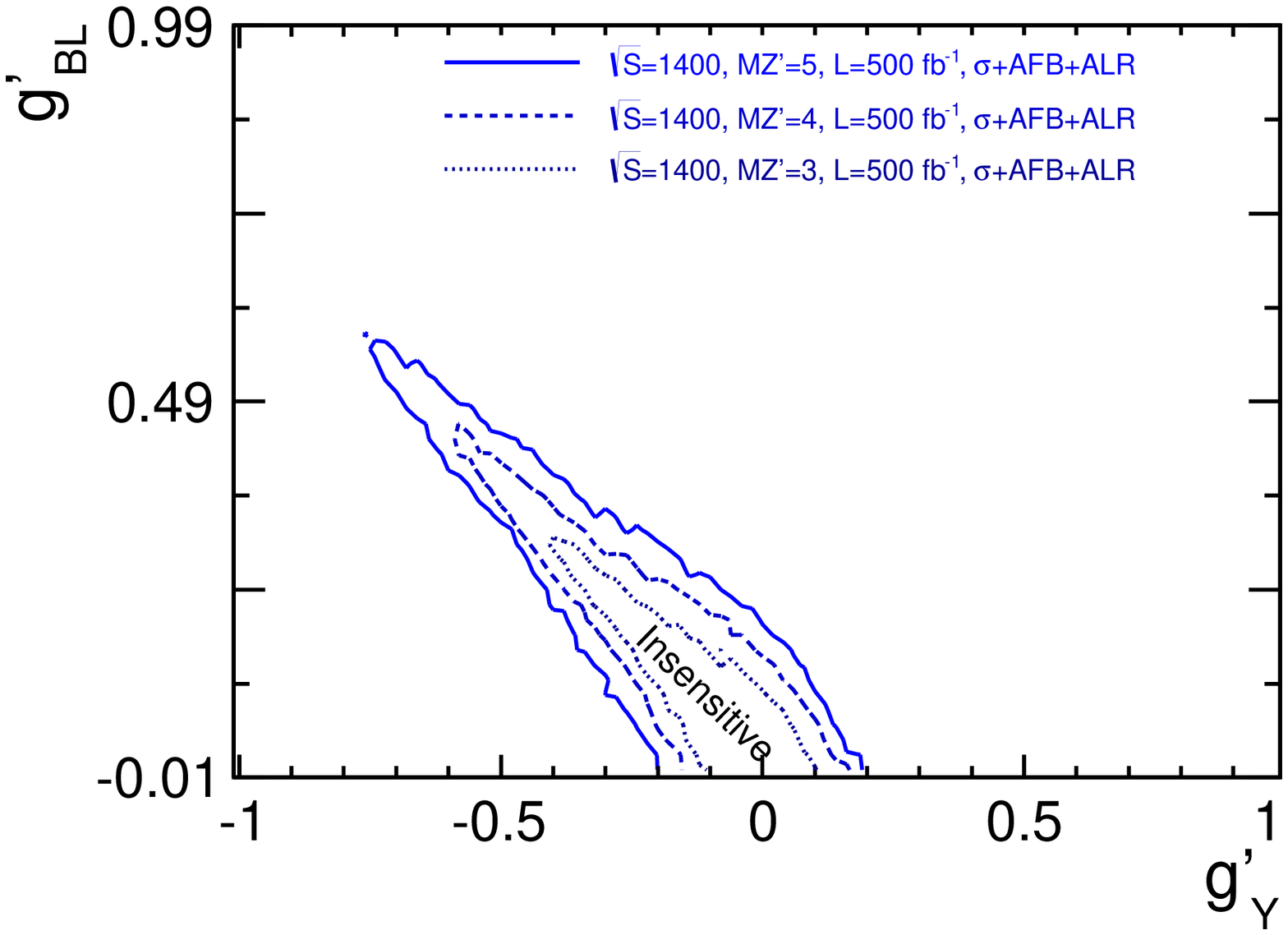} }
\end{tabular}
}
\end{center}
\caption{$5\sigma$ discovery potential in ($g'_Y,g'_{BL}$) plane, determined from combined observables 
$\sigma$+$A_{FB}$+$A_{LR}$ at $\sqrt{s}=1.4\tev$ for
(a) $\mzp=5\tev$ and different luminosities,
(b) L=500 $\mathrm{fb^{-1}}$ and different $\mzp$ values (same as Figure~\ref{fig:DISCP3:1400} except $A_{LR}$ added ).}
\label{fig:DISCP4:1400}
\end{figure}

Figure~\ref{fig:DISCP4:1400} shows the $5\sigma$ discovery potential in the ($g'_Y,g'_{BL}$) plane,  
determined from the combined observables $\sigma$ + $A_{FB}$ +  $A_{LR}$, at $1.4\tev$, 
(a) $\mzp=5\tev$ and different luminosity values, 
(b) L=500 $\mathrm{fb^{-1}}$ and different $\mzp$ values.

\begin{figure}[tp]
\begin{center}
\resizebox{\textwidth}{!} {
\begin{tabular}{c}
\hspace{-1.cm}
\subfloat[$\mzp=6\tev$, L=250, 500 and 1000 $\mathrm{fb^{-1}}$ ]
{\includegraphics[width=0.49\textwidth,clip]{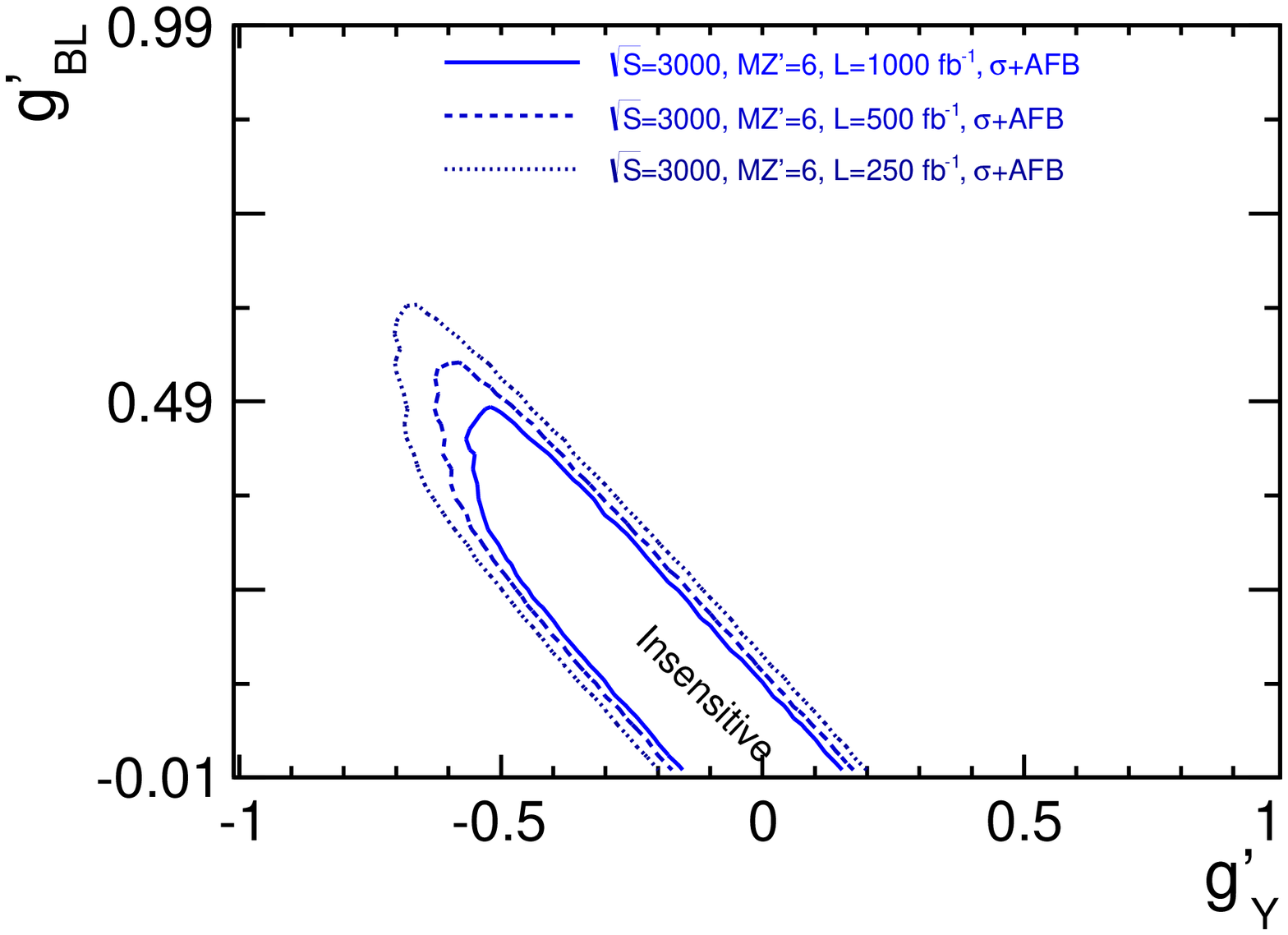}}  
\subfloat[L=500 $\mathrm{fb^{-1}}$  $\mzp=4, 7, 10\tev$]
{\includegraphics[width=0.49\textwidth,clip]{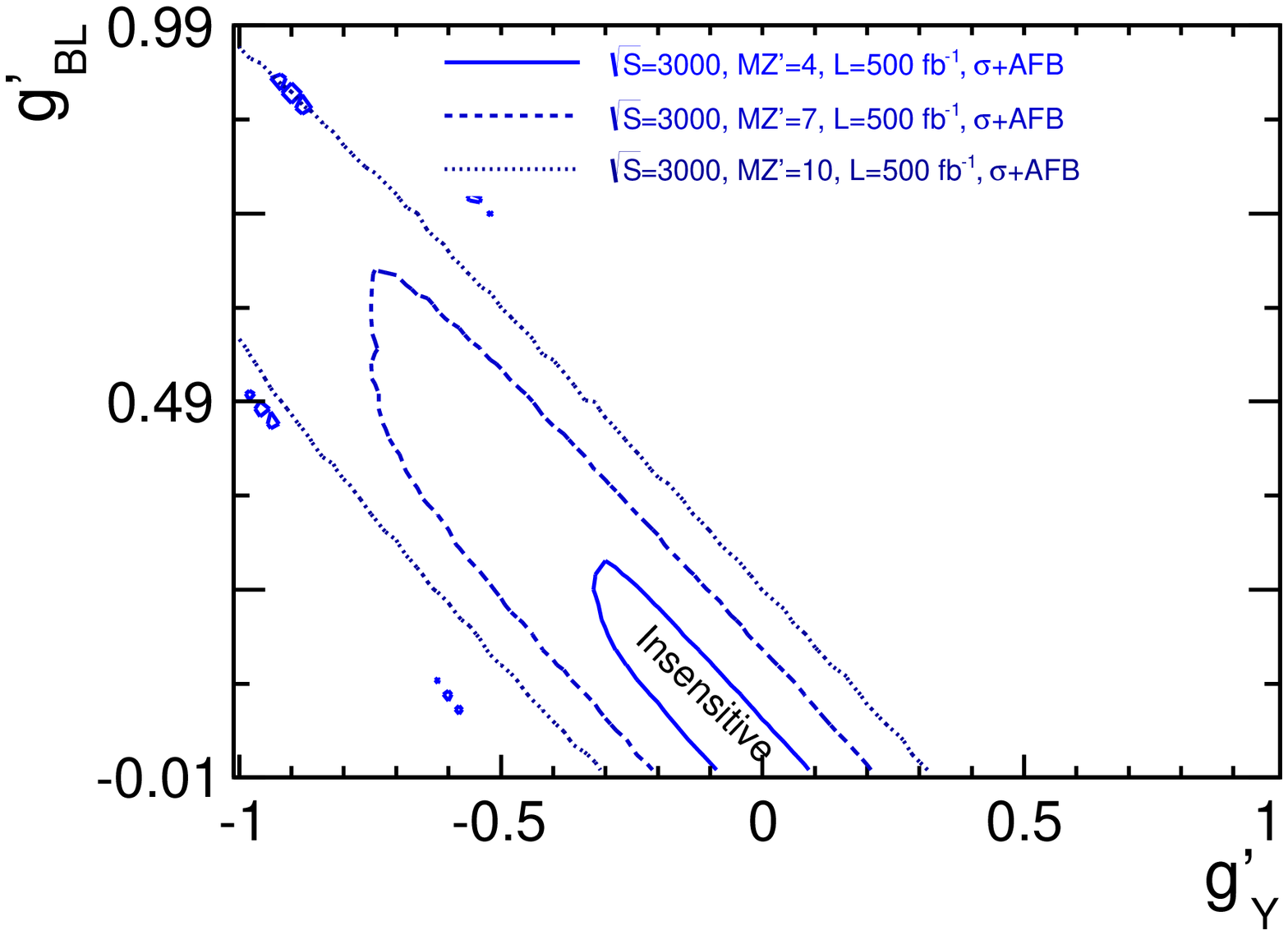} }
\end{tabular}
}
\end{center}
\caption{$5\sigma$ discovery potential in ($g'_Y,g'_{BL}$) plane, determined from combined observables 
$\sigma$+$A_{FB}$ at $\sqrt{s}=3\tev$ for (a) $\mzp=6\tev$ and different luminosities,
(b) L=500 $\mathrm{fb^{-1}}$ and different $M_{Z'}$ values }
\label{fig:DISCP1:3000}
\end{figure}

Figure~\ref{fig:DISCP1:3000} shows the $5\sigma$ discovery potential in the ($g'_Y,g'_{BL}$) plane,  determined 
from the combined observables $\sigma$ + $A_{FB}$, at $3\tev$, 
(a) $\mzp=6\tev$ and different luminosity values, 
(b) L=500 $\mathrm{fb^{-1}}$ and different $\mzp$ values.

\begin{figure}[h]
\begin{center}
\resizebox{\textwidth}{!} {
\begin{tabular}{c}
\hspace{-1.cm}
\subfloat[$\mzp=6\tev$, L=250, 500 and 1000 $\mathrm{fb^{-1}}$ ]
{\includegraphics[width=0.50\textwidth,clip]{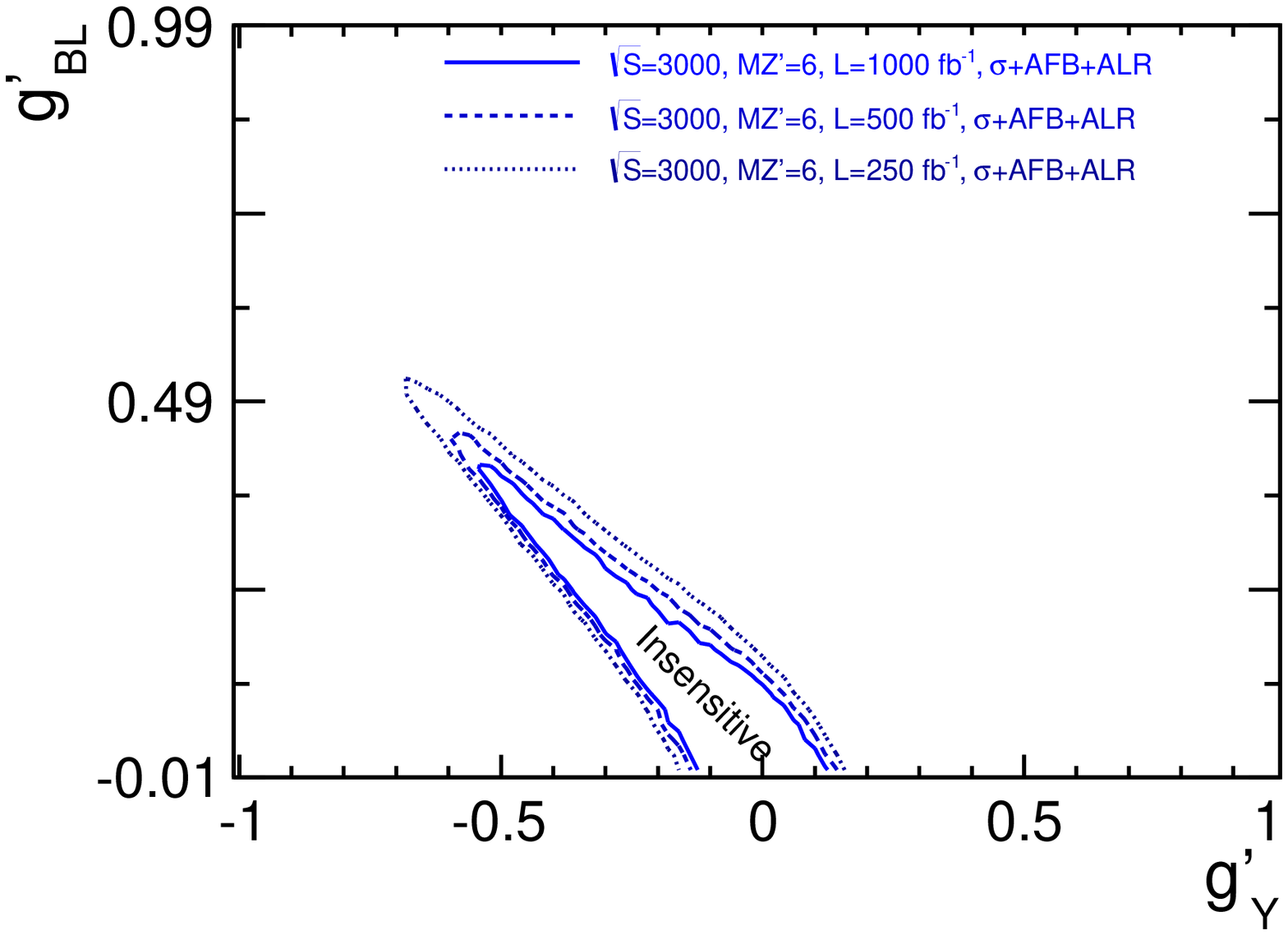}}  
\subfloat[L=1000 $\mathrm{fb^{-1}}$  $M_{Z'}$=4, 7, 10 TeV]
{\includegraphics[width=0.50\textwidth,clip]{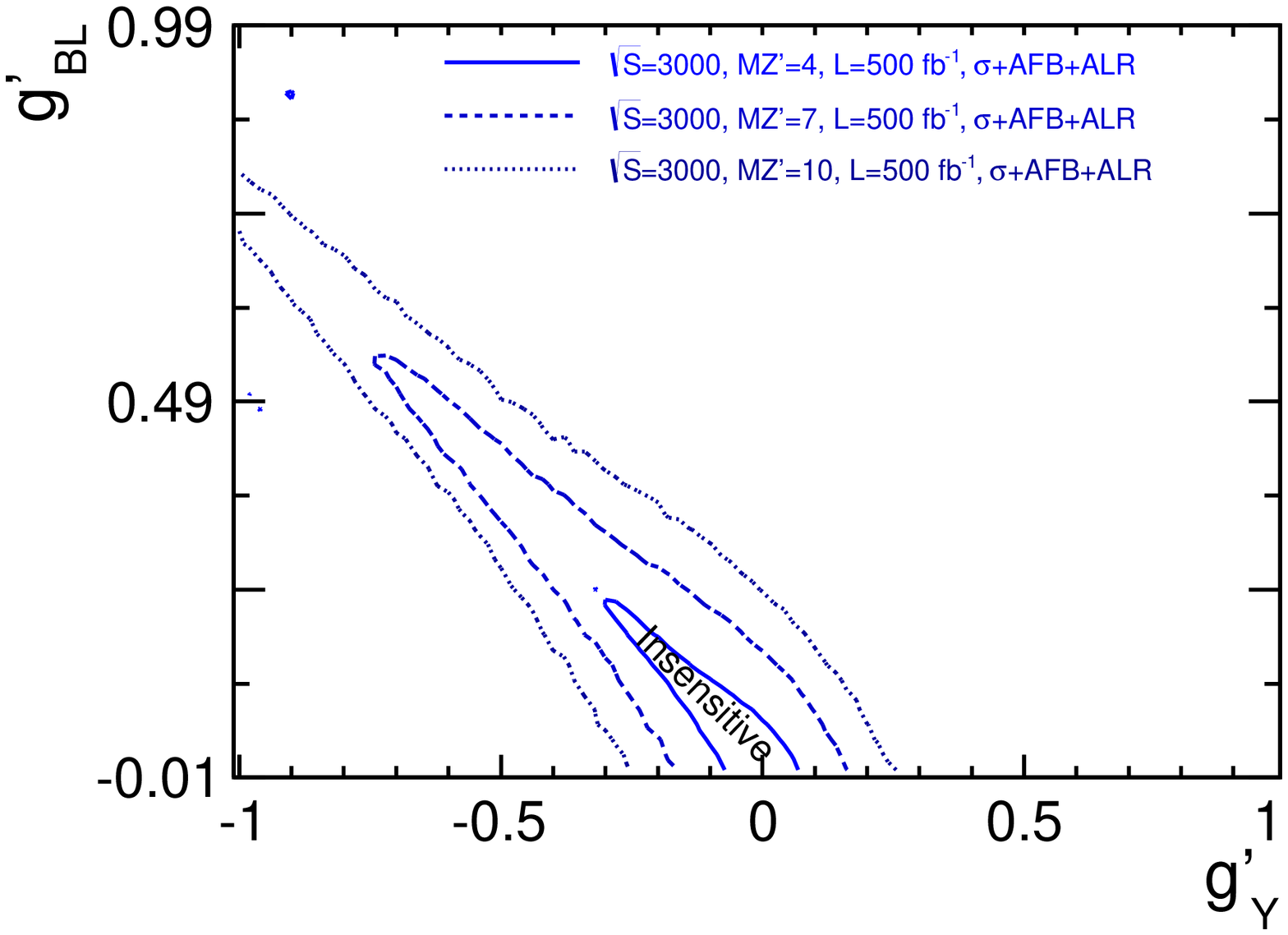} }
\end{tabular}
}
\end{center}
\caption{$5\sigma$ discovery potential in ($g'_Y,g'_{BL}$) plane, determined from combined observables 
$\sigma$+$A_{FB}$+$A_{LR}$ at 
$\sqrt{s}=3\tev$ for
(a) $\mzp=6\tev$ and different luminosities,
(b) L=500 $\mathrm{fb^{-1}}$ and different $\mzp$ values (same as Figure~\ref{fig:DISCP1:3000} except $A_{LR}$ added ).}
\label{fig:DISCP2:3000}
\end{figure}

Figure~\ref{fig:DISCP2:3000} shows the $5\sigma$ discovery potential in the ($g'_Y,g'_{BL}$) plane,  
determined from the combined observables $\sigma$ + $A_{FB}$ +  $A_{LR}$, at $3\tev$, 
(a) $\mzp=6\tev$ and different luminosity values, 
(b) L=500 $\mathrm{fb^{-1}}$ and different $\mzp$ values

\begin{figure}[h]
\begin{center}
\resizebox{\textwidth}{!} {
\begin{tabular}{c}
\hspace{-1.cm}
%\subfloat[MZ'=6 TeV, L=250, 500 and 1000 $\mathrm{fb^{-1}}$ ]
{\includegraphics[width=1.0\textwidth,clip]{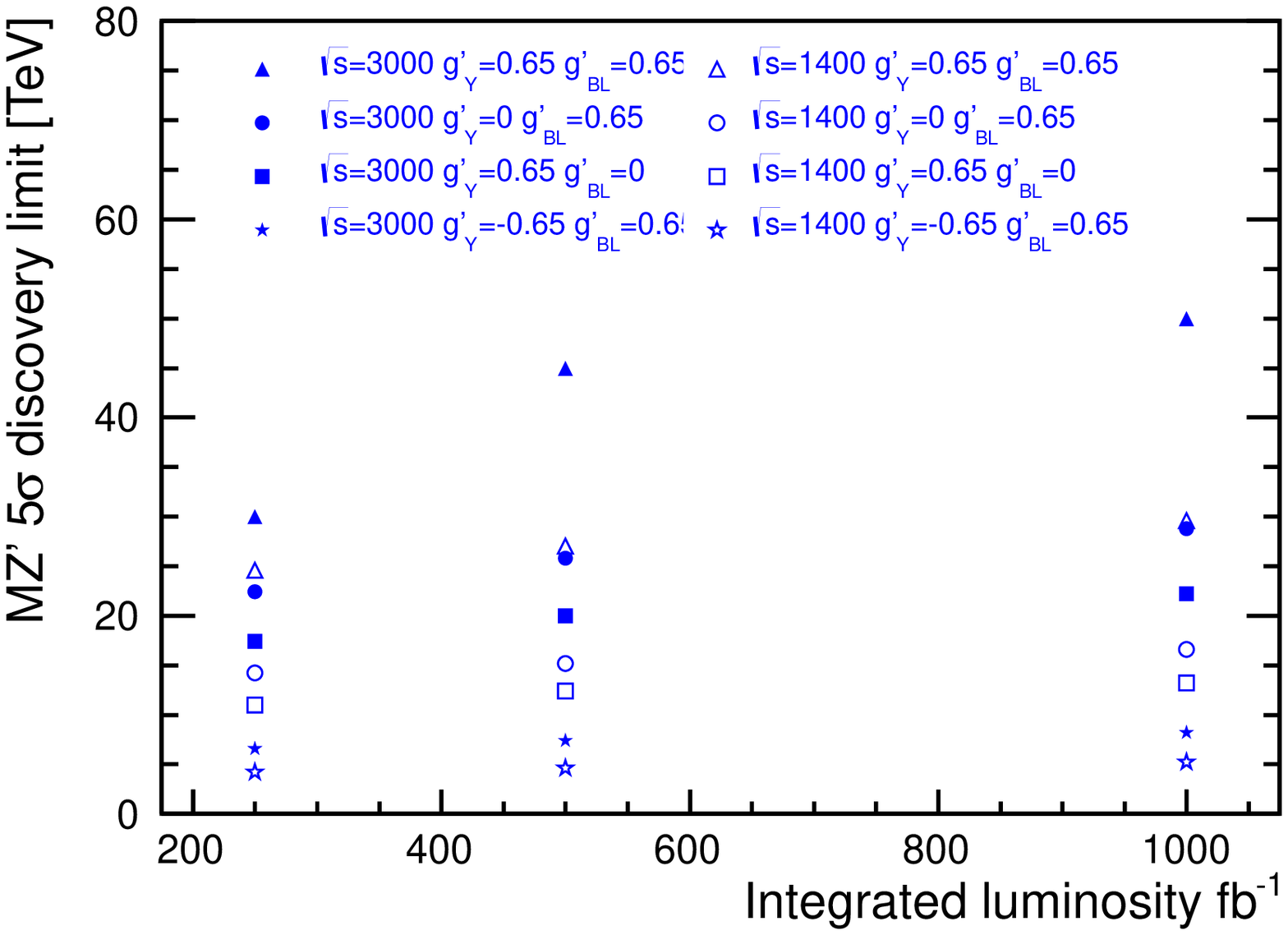}}  
\end{tabular}
}
\end{center}
\caption{$\mzp$ $5\sigma$ discovery limit as function of the
integrated luminosity for different values of the couplings $g'_Y$ and $g'_{BL}$.
The limits shown are determined from the combined observables $\sigma$ + $A_{FB}$ at $3\tev$ and $1.4\tev$.
}
\label{fig:DISCPM:3000:1400}
\end{figure}

Figure~\ref{fig:DISCPM:3000:1400} shows the $\mzp$ $5\sigma$ discovery limit, as function of the
integrated luminosity for different values of the couplings $g'_Y$ and $g'_{BL}$.  
The limits shown are determined from the combined observables $\sigma$ + $A_{FB}$, at $3\tev$ and $1.4\tev$. 
For negative values of $g'_Y$ the limits are significantly lower.  As a check we have applied our methodologies to LEP2 energy and integrated luminosities and compared $3\sigma$ exclusion limits of the $B-L$ model to that obtained by~\cite{Carena:2004xs} and find good agreement.

%%%%%%%%%%%%%%%%%%%%%%%%%%%%%%%%%%%%%%%%%%%%%
\section{Model-Dependent Couplings Determination} \label{dtsmcs again}

Assuming LHC discovers a $Z'$ of mass 5 TeV, the couplings can be determined making a model assumption.
The $\mafz$ predictions of the observables $\sigma(\mafz)$, $A_{FB}(\mafz)$ and $A_{LR}(\mafz)$
are computed for $\mzp=5\tev$ and for different values of $g'_Y$ and $g'_{BL}$.
For each observable the $\chi^2$ is computed. 
\begin{eqnarray}
\resizebox{0.9\textwidth}{!} {
$\chi_{\sigma}^2=\frac{(\sigma(\mathrm{\mafz})-\sigma(\mathrm{Data}))^2}{\Delta\sigma(\mathrm{Data})^2}, $ \hspace{0.25cm}
$\chi_{A_{FB}}^2=\frac{(A_{FB}(\mathrm{\mafz})-A_{FB}(\mathrm{Data}))^2}{ \Delta A_{FB}(\mathrm{Data})^2}, $ \hspace{0.25cm}
$\chi_{A_{LR}}^2=\frac{(A_{LR}(\mathrm{\mafz})-A_{LR}(\mathrm{Data}))^2}{\Delta A_{LR}(\mathrm{Data})^2} $
}
\end{eqnarray}
where $  \Delta\sigma(\mathrm{Data}),~ \Delta A_{FB}(\mathrm{Data})~ \mathrm {and}~ \Delta A_{LR}(\mathrm{Data})$ 
are the experimental errors on the measurement of the observables in the presence of $Z'$ of mass $5\tev$.  To determine the couplings,
$\chi_{\sigma}^2,~ \chi_{A_{FB}}^2, ~\chi_{A_{LR}}^2$ and $\chi_{sum}^2$=$\chi_{\sigma}^2+\chi_{A_{FB}}^2+\chi_{A_{LR}}^2$ are computed for different values of $g'_Y$ and $g'_{BL}$.
The polarization value and the systematic errors are the same as in the previous section. 
The model chosen is tested for compatibility with the data by determining if it has a 
sufficiently low minimal $\chi^2_{sum}$. 

\begin{figure}[htbp]
\begin{center}
\resizebox{\textwidth}{!} {
\begin{tabular}{c}
\hspace{-1.cm}
\subfloat[$\sigma$+$A_{FB}$+$A_{LR}$ ]
{\includegraphics[width=0.50\textwidth,clip]{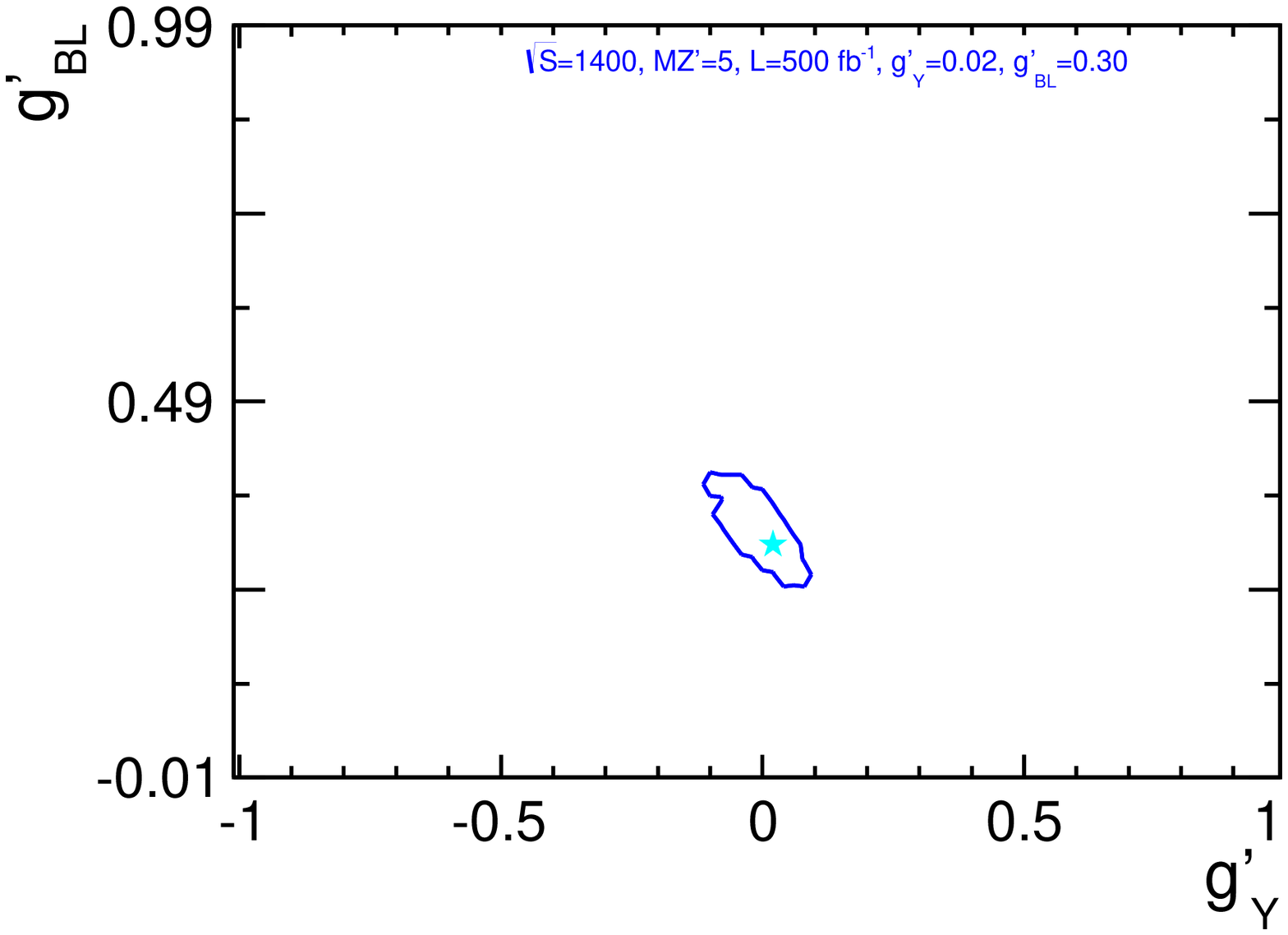}}  
\subfloat[$\sigma$+$A_{FB}$]
{\includegraphics[width=0.50\textwidth,clip]{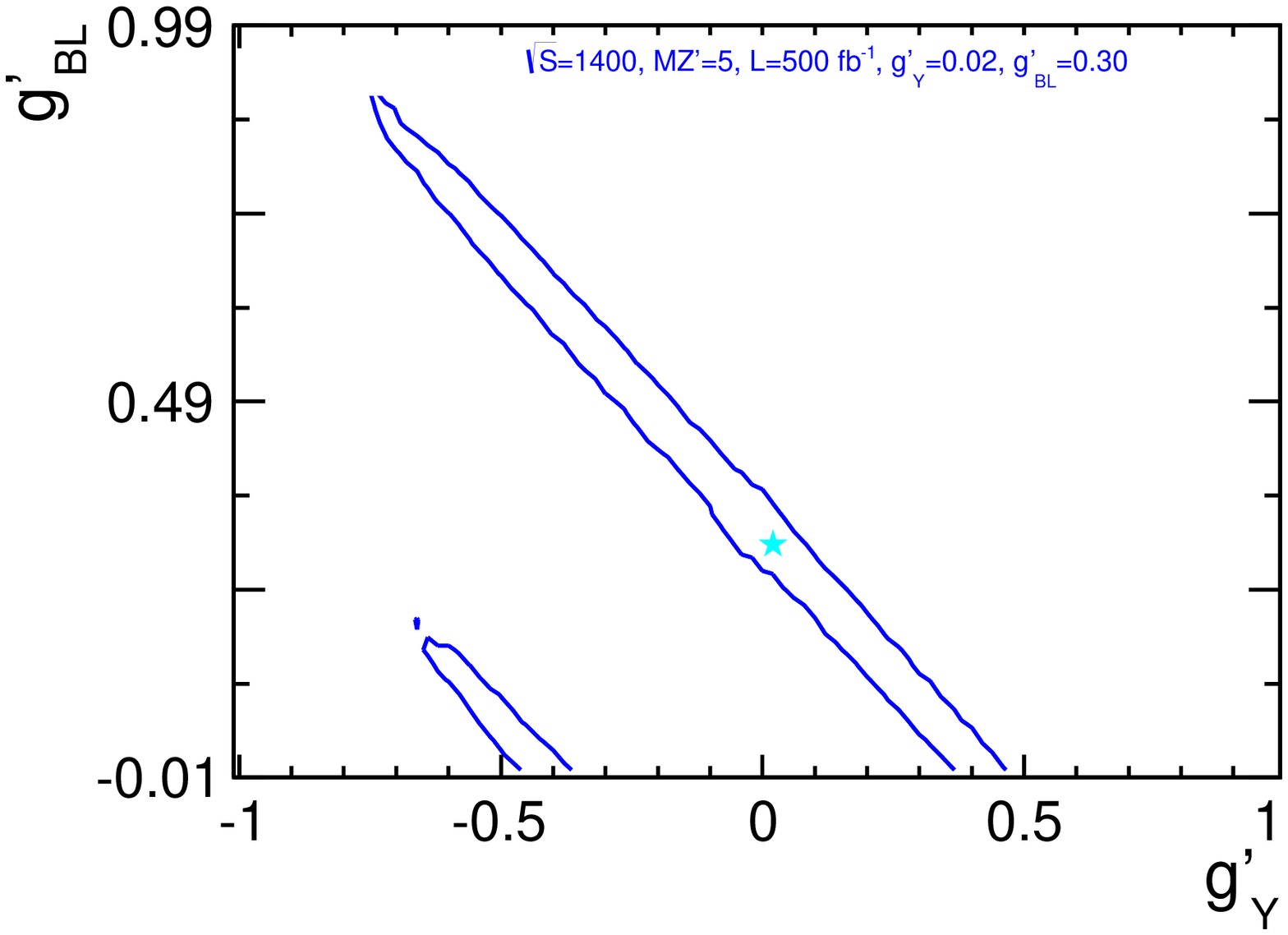} }
\end{tabular}
}
\end{center}
\caption{$3\sigma$ couplings contour in ($g'_Y,g'_{BL}$) plane, determined from combined observables 
(a)$\sigma$+$A_{FB}$+$A_{LR}$,
(b)$\sigma$+$A_{FB}$,
$\sqrt{s}=1.4\tev$, $\mzp=5\tev$ and L=500 $\mathrm{fb^{-1}}$  $g'_Y=0.02$ and $g'_{BL}=0.3$}
\label{fig:COUP1:1400}
\end{figure}

Figure~\ref{fig:COUP1:1400} shows the $3\sigma$ contour in the ($g'_Y,g'_{BL}$) plane for $\mzp=5\tev$, $\sqrt{s}=1.4\tev$, 
L=500 $\mathrm{fb^{-1}}$ , $g_Y=0.02$ and $g_{BL}=0.3$,
determined from the combined observables, 
(a) $\sigma$ + $A_{FB}$+$A_{LR}$,
(b) $\sigma$ + $A_{FB}$.
It shows that without the  $A_{LR}$ observable, whose measurement is made possible by polarized electron beam, 
the couplings could not be determined.

\begin{figure}[htbp]
\begin{center}
\resizebox{\textwidth}{!} {
\begin{tabular}{c}
\hspace{-1.cm}
%%\vspace{-1.0cm}
\subfloat[ $g'_Y=-0.5$ and $g'_{BL}=0.02$ ]
{\includegraphics[width=0.50\textwidth,clip]{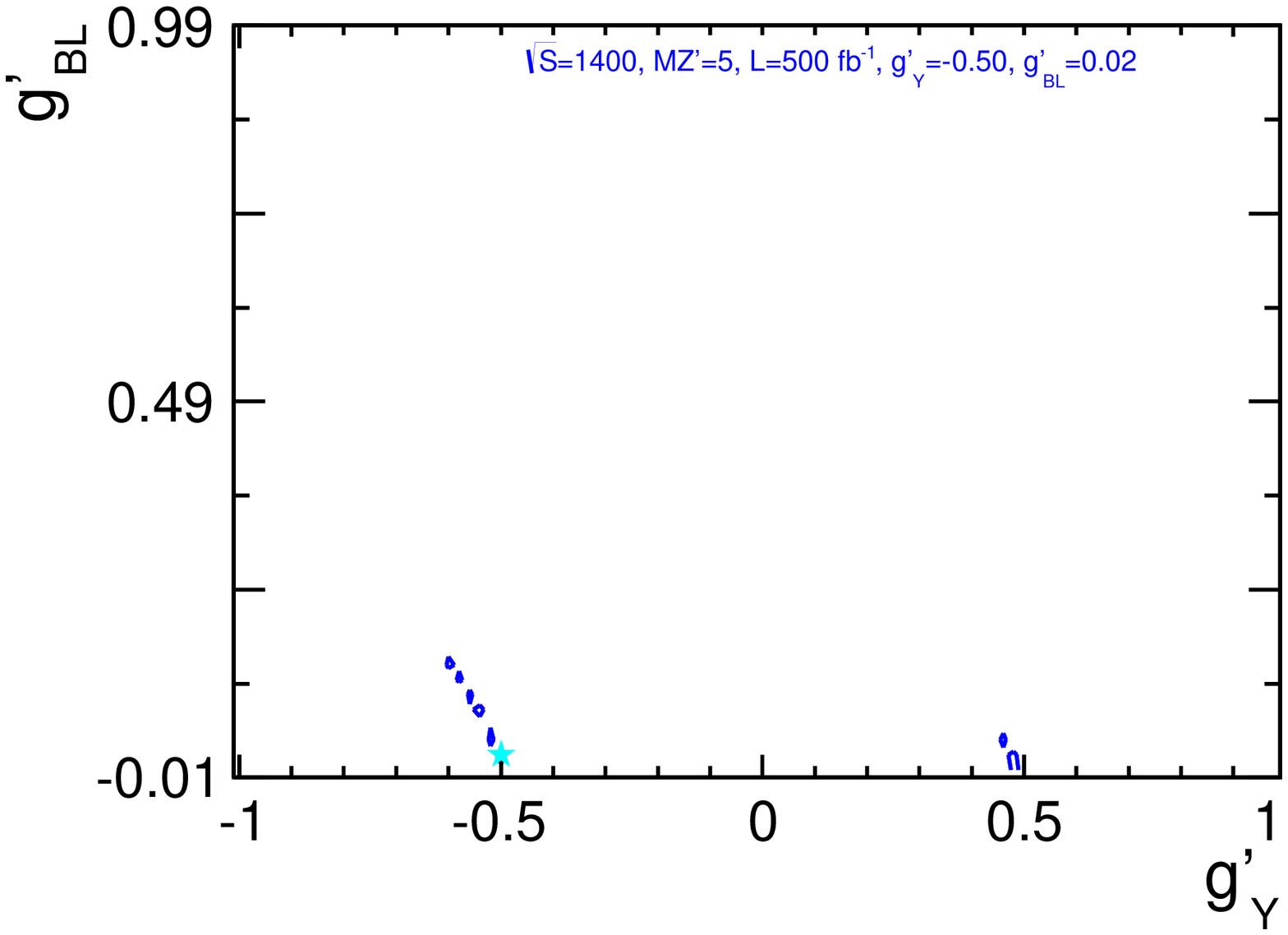}}  
\subfloat[  $g'_Y=0.5$ and $g'_{BL}=0.02$]
{\includegraphics[width=0.50\textwidth,clip]{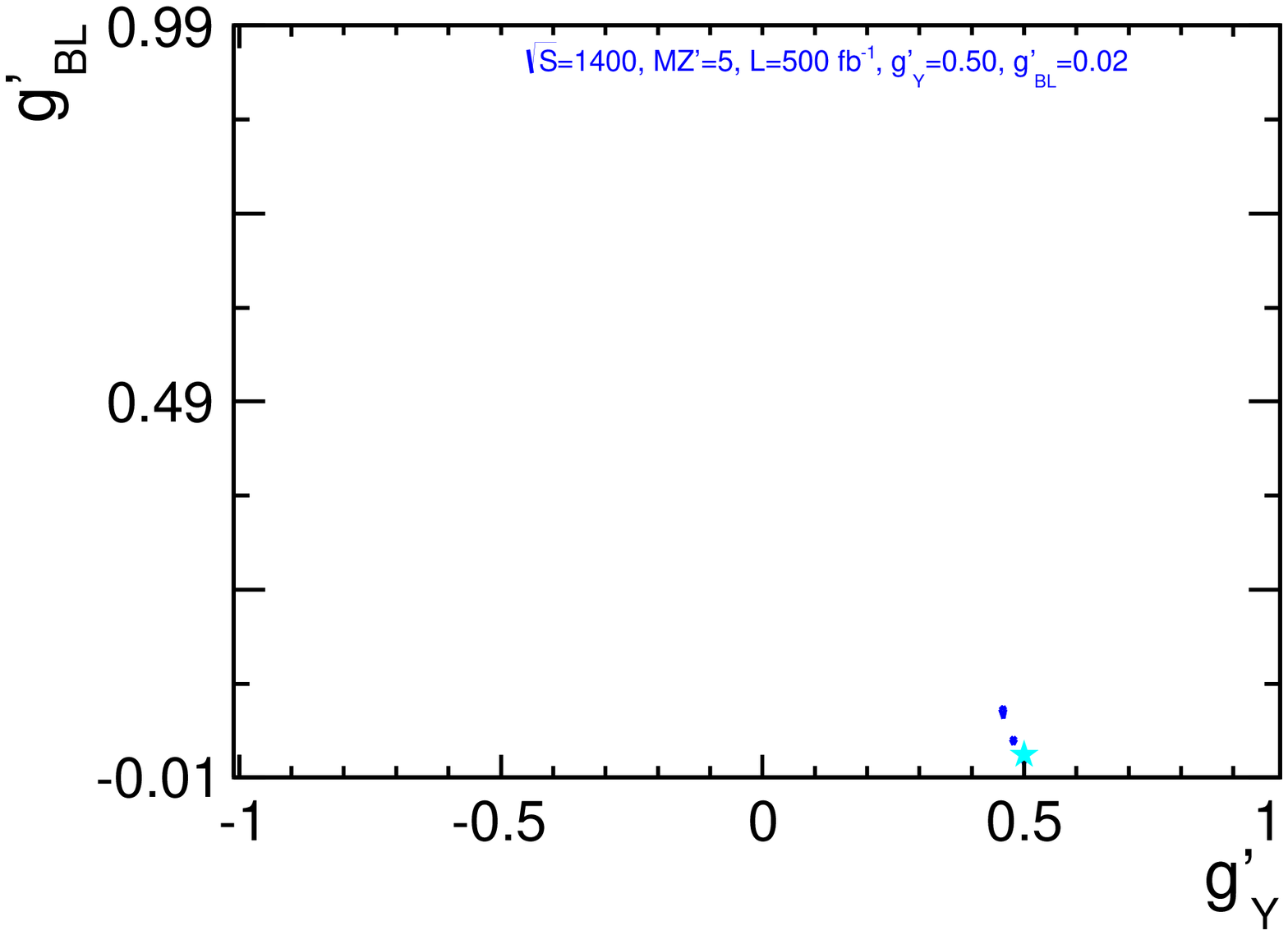} }
\end{tabular}
}
\end{center}
%\vspace{-0.5cm}
\caption{$3\sigma$ couplings contour in ($g'_Y,g'_{BL}$) plane, determined from combined observables $\sigma$+$A_{FB}$+$A_{LR}$,
(a) $g'_Y=-0.5$ and $g'_{BL}=0.02$,
(b) $g'_Y=0.5$ and $g'_{BL}=0.02$,
$\sqrt{s}=1.4\tev$, $\mzp=5\tev$ 
and L=500 $\mathrm{fb^{-1}}$  }
\label{fig:COUP2:1400}
\end{figure}

Figure~\ref{fig:COUP2:1400} shows the 3$\sigma$ contour in the ($g'_Y,g'_{BL}$) plane 
determined from the combined observables, $\sigma$ + $A_{FB}$+ $A_{LR}$
for $\mzp=5\tev$, $\sqrt{s}=1.4\tev$, L=500 $\mathrm{fb^{-1}}$, 
(a) $g'_Y=-0.5$ and $g'_{BL}=0.02$, (b) $g'_Y=0.5$ and $g'_{BL}=0.02$.
It shows that for low values of $g'_{BL}$ and negative values of $g'_{Y}$ two solutions
can be found.

\begin{figure}[htbp]
\begin{center}
\resizebox{\textwidth}{!} {
\begin{tabular}{c}
\hspace{-1.cm}
%\vspace{-1.cm}
\subfloat[ $g'_Y=0.02$ and $g'_{BL}=0.2$ ]
{\includegraphics[width=0.50\textwidth,clip]{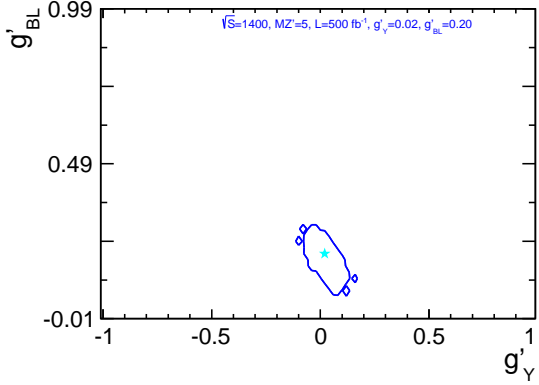}}  
\subfloat[  $g'_Y=0.02$ and $g'_{BL}=0.1$]
{\includegraphics[width=0.50\textwidth,clip]{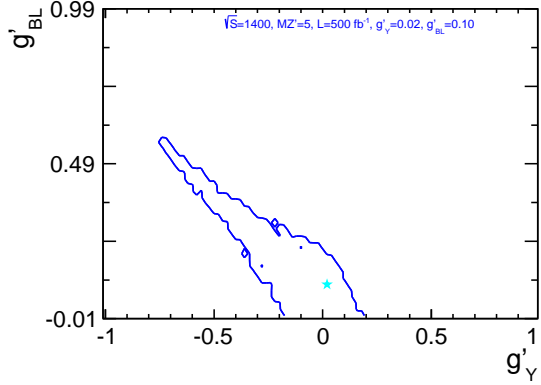} }
\end{tabular}
}
\end{center}
\caption{$3\sigma$ couplings contour in ($g'_Y,g'_{BL}$) plane, determined from combined observables $\sigma$+$A_{FB}$+$A_{LR}$,
(a) $g'_Y=0.02$ and $g'_{BL}=0.2$,
(b) $g'_Y=0.02$ and $g'_{BL}=0.1$,
$\sqrt{s}=1.4\tev$, $\mzp=5\tev$ and L=500 $\mathrm{fb^{-1}}$  }
\label{fig:COUP3:1400}
\end{figure}

Figure~\ref{fig:COUP3:1400} shows the $3\sigma$ contour in the ($g'_Y,g'_{BL}$) plane 
determined from the combined observables, $\sigma$ + $A_{FB}$+ $A_{LR}$
for $\mzp=5\tev$, $\sqrt{s}=1.4\tev$, L=500 $\mathrm{fb^{-1}}$.
(a) $g'_Y=0.02$ and $g'_{BL}=0.2$, (b) $g'_Y=0.02$ and $g'_{BL}=0.1$.
It shows that for low values of $g'_{Y}$ and low values of $g'_{BL}$ the error on
the couplings can be very large.
 
\ifnote
\clearpage
\fi

%%%%%%%%%%%%%%%%%%%%%%%%%%%%%%%%%%%%%%%%%
\section{Summary}

The $Z'$ discovery potential and the accuracies in the determination of the $Z' \mu^+\mu^-$ couplings have been studied at CLIC at 1.4 and $3\tev$ in the framework of the $\mafz$ model.
The analysis is based on dimuon events for which the SM background
processes and the beam-induced background can be removed by selection cuts. 
The signal selection efficiency is 5.9\% at $3\tev$ and 15.1\% at $1.4\tev$.
While polarized beams give only a small improvement to the discovery potential, they are essential
for the determination of the $Z'\mu^+\mu^-$ couplings.

Assuming the LHC discovers a $Z'$ it will likely be through resonance signal of a $Z'$ with mass less 
than $\sim 6\tev$~\cite{OtherZprimestudies}. Let us assume that a discovery is made at the LHC of a $Z'$ mass 
peak at $5\tev$. In that case one of the free parameters will be determined, and from our CLIC observables we 
first can determine if the $\mafz$ is consistent with the data, and if yes, can pin down the 
couplings $g'_Y$ and $g'_{BL}$. How well CLIC will be able to pin down these couplings depends on precisely what 
values they have. Over the majority of parameter space illustrated in this work, these couplings can be 
determined to within $2-20\%$. The lower value, $2\%$, qualifies for $g'_Y$ and $g'_{BL}$ couplings both 
being positive. The upper value, $20\%$, qualifies for $g'_Y$ negative and $g'_{BL}$ positive, for example.  
Thus, as is the case in all beyond-the-SM theories, the sensitivities to the new theory are determined by the 
details of the theory, i.e., the values of its couplings. Nevertheless, the sensitivities are impressive 
throughout the parameter space of $\mafz$, except when both couplings are small, 
and complement well the capabilities of the LHC to find the resonance. 

If the $Z'$ state is too heavy to be found at the LHC, the theory is unlikely to cause any deviation at all in 
LHC observables. On the other hand, CLIC observables can register a clear deviation away from the 
SM even if $\mzp$ is well above the center of mass energy of the machine. ``Reduced couplings" that include 
unknown $\mzp$ factors in them can be determined~\cite{Leike:1996pj}. 
For example, one can see deviations from the 
SM with $g'_Y=g'_{BL}=0.65(=g_2$ of the SM) for $\mzp$ mass values up to $30\tev$ for a $1.4\tev$ collider, 
and up to $50\tev$ for a $3\tev$ collider. 
This excellent mass reach is a well-known positive feature of the 
physics potential of an $e^+e^-$ collider, and this study demonstrates this straightforwardly within the simple 
case of the $\mafz$ model. 

\ifjhep
%\clearpage
\fi

\section*{Acknowledgments}
We are grateful to J. Reuter for implementing the $\mafz$ model in Whizard, and to G. Villadoro for discussions.

\end{document}